\documentclass[twocolumn]{aastex62}
\usepackage[figuresright]{rotating}
\usepackage{multirow}
\usepackage{array}
\usepackage{amsmath,amstext}
\usepackage{CJK}
\usepackage[T1]{fontenc}
\usepackage[figure,figure*]{hypcap}
\usepackage{booktabs}
\usepackage{cleveref}
\usepackage{paralist}
\usepackage{natbib}
\usepackage{graphicx}
\usepackage{hyperref} 
\graphicspath{{fig/}}

\usepackage{threeparttable}

\newcommand{\kms}{km\,$\rm s^{-1}$}
\newcommand{\masyr}{\,mas\,$\rm yr^{-1}$}

\newcommand {\Msun}{{M$_{\odot}$}}

\usepackage{pgfplotstable} 
\usepackage{booktabs}  
\usepackage{lipsum} 

\shorttitle{Multiple generation star formation in Cepheus Flare}
\shortauthors{Wang et al.}
\usepackage{lineno}

\begin{document}
\begin{CJK*}{UTF8}{gbsn}

\title{Multiple generation star formation in Cepheus Flare}

\correspondingauthor{Min Fang}
\email{mfang@pmo.ac.cn}

\author{Fan Wang}
\affil{Purple Mountain Observatory, Chinese Academy of Sciences, Nanjing 210023, China}
\affil{School of Astronomy and Space Science, University of Science and Technology of China, Hefei, Anhui 230026, China}

\author{Min Fang}
\affil{Purple Mountain Observatory, Chinese Academy of Sciences, Nanjing 210023, China}
\affil{School of Astronomy and Space Science, University of Science and Technology of China, Hefei, Anhui 230026, China}

\author[0000-0002-6506-1985]{Xiaoting Fu}
\affil{Purple Mountain Observatory, Chinese Academy of Sciences, Nanjing 210023, China}
\affil{School of Astronomy and Space Science, University of Science and Technology of China, Hefei, Anhui 230026, China}

\author{Yang Su}
\affil{Purple Mountain Observatory, Chinese Academy of Sciences, Nanjing 210023, China}
\affil{School of Astronomy and Space Science, University of Science and Technology of China, Hefei, Anhui 230026, China}

\author{Xuepeng Chen}
\affil{Purple Mountain Observatory, Chinese Academy of Sciences, Nanjing 210023, China}
\affil{School of Astronomy and Space Science, University of Science and Technology of China, Hefei, Anhui 230026, China}

\author{Shiyu Zhang}
\affil{Purple Mountain Observatory, Chinese Academy of Sciences, Nanjing 210023, China}
\affil{School of Astronomy and Space Science, University of Science and Technology of China, Hefei, Anhui 230026, China}

\author{Penghui Liu}
\affil{Purple Mountain Observatory, Chinese Academy of Sciences, Nanjing 210023, China}
\affil{School of Astronomy and Space Science, University of Science and Technology of China, Hefei, Anhui 230026, China}

\author[0000-0003-2536-3142]{Xiao-Long Wang}
\altaffiliation{Physics Postdoctoral Research Station at Hebei Normal University.}
\affiliation{Department of Physics, Hebei Normal University, Shijiazhuang 050024, People's Republic of China}
\affiliation{Guo Shoujing Institute for Astronomy, Hebei Normal University, Shijiazhuang 050024, People's Republic of China}
\affiliation{Hebei Advanced Thin Films Laboratory, Shijiazhuang 050024, People's Republic of China}

\author{Haijun Tian}
\affiliation{School of Science, Hangzhou Dianzi University, Hangzhou 310018, China} 
\affiliation{Zhejiang Branch of National Astronomical Data Center, Hangzhou 310018, China}

\author{Wenyuan Cui}
\affiliation{College of Physics, Hebei Normal University, Shijiazhuang 050024, China}

\author{Zhongmu Li}
\affiliation{Institute of Astronomy and information, Dali University, Dali 671003, China}


\begin{abstract}
We present an analysis of the young stellar moving group ASCC~127 using Gaia~DR3 data, significantly expanding its membership to 3,971 stars -- double the number identified in previous studies. Using kinematic and distance criteria, ASCC~127 is divided into five subgroups (Groups~1$-$5) with ages spanning from 15 to 32~Myr. Groups~1$-$5 are spatially linked to the Cepheus Flare star-forming region, revealing potential evidence of four sequential star formation episodes at approximately 32~Myr, 20~Myr, 15~Myr, and 7~Myr. Through dust and gas mapping, we identify a spatial cavity extending several tens of parsecs, which may have resulted from feedback processes such as supernovae associated with earlier generations of stars in the region. This structure, along with the larger Loop~III feature, indicates that feedback from massive stars likely influenced the interstellar medium (ISM). By integrating young stellar populations with ISM studies, we provide a detailed picture of the feedback-driven star formation history in the Cepheus Flare region.
\end{abstract}



\section{Introduction}
\label{sec:intro}

Star formation is a fundamental process in galactic evolution, involving complex dynamics, radiation, and chemical interactions within the interstellar medium \citep[ISM, see e.g.,][]{Schinnerer2024,Kachelriess2024}. In the Milky Way, star formation typically occurs within molecular clouds, where high-density gas environments are clearly influenced by stellar feedback mechanisms. Observational and theoretical studies increasingly reveal that star formation often proceeds in sequential episodes, influenced by feedback from massive stars. This feedback, in the form of stellar winds, radiation, and supernova explosions, impacts the surrounding ISM, compressing molecular clouds and potentially triggering the collapse of new stellar populations \citep[e.g.,][]{Zari2019,Esplin2022,Zhou2022,Zucker2023,Sanchez2024}. Understanding these multi-generational processes is one of the keys for reconstructing the star formation history of complex regions in the Milky Way.

In theoretical models, numerous studies have employed numerical simulations to investigate how stellar feedback forms complex structures in the interstellar medium and accelerates star formation \citep[e.g.,][]{Grudic2021a, Grudic2021b, Grudic2022, Guszejnov2022}. These studies suggest that supernova explosions and intense stellar winds can clear the surrounding gas while also compressing nearby molecular clouds, thereby increasing local density and fostering conditions favorable to the formation of subsequent stellar generations. In regions of lower density, feedback can also produce visually observable cavities, further influencing the evolution of the interstellar medium \citep[e.g.,][]{Larson1971,Kim2017,Grudic2022}.

In recent years, observational evidence for multi-generational star formation has increased, particularly with the release of Gaia data. Gaia's extensive high-precision astronomic and photometric data provide essential support for examining the mechanisms and history of multi-generational star formation within star-forming regions \citep[e.g.,][]{Zari2019,Esplin2022,Zhou2022,Zucker2023,Sanchez2024}. Analyzing the positions, age distribution, and kinematics of stars can reveal sequential star formation process, which are crucial for understanding the evolution of stars and stellar groups.

In particular, for star-forming regions near the Sun, young (1 to 30~Myr) moving groups play an essential role in uncovering the mechanisms and history of multi-generation star formation. Nearby stellar groups such as the Orion association and the Scorpius-Centaurus association provide observational evidence that often shows age gradients across different groups or clusters. These gradients suggest a sequentially triggered star formation process. For instance, in the Scorpius-Centaurus association, observational evidence indicates that stellar subgroups of varying ages are spatially distributions of massive stars triggering star formation at greater distances \citep[e.g.,][]{MiretRoi2022, Briceno2023, Ratzenbock2023}.

Many such young, nearby moving groups have been found. For example, the catalog provided by \citet{Kounkel(2019)} includes hundreds of these moving groups, one of which is Theia~57, (ASCC~127), which has a relatively complex structure. It is a young stellar moving group situated near the Cepheus Flare star-forming region. Previous studies identified ASCC~127 as a loose association of stars spanning a distance range of 300$-$500~pc. However, its full membership, internal structure, and relationship with the Cepheus Flare have remained unclear. Using the high-precision astrometric and photometric data from Gaia~DR3, we  re-identified ASCC~127 with the coherent algorithm (i.e., Friends-of-Friends, FoF), significantly expanding its membership and uncovering its complete structure. We depict the comprehensive properties (e.g., kinematics, ages and mass) of ASCC~127. The structures of ASCC~127, along with the young stellar populations in the Cepheus Flare, provide critical clues to the region’s star formation history.

We organize the paper as follows. In Section~\ref{sec:Data} we briefly describe the data used in this work, and Section~\ref{sec:membership} show the target selection and membership and the ages and masses. Finally, we provide a discussion and our summary in Sections~\ref{sec:discu} and \ref{sec:summary}, respectively.

\section{Data and model}
\label{sec:Data}
We integrate data from Gaia~DR3 star sources, interstellar dust maps, CO maps, and HI maps to investigate the spatial, kinematic, and ISM properties of ASCC~127 and its surrounding environment, and also briefly introduce the PARSEC 1.2S isochrone model we use in this work. In the following, we will briefly introduce these data sources.

\subsection{Gaia~DR3}
\label{sec:gaiadr3}
In this work, we use 5D phase-space information, i.e., $\ell$, b, $\mu_{\ell^*}$, $\mu_b$ and distances, to uncover the member candidates from the third release of the Gaia Data  \citep[DR3, ][]{Gaia2022}. Gaia~DR3 provides astrometric information with the high-precision for about 1.8 billion sources over the sky and near-mmag precision photometric data in BP, RP and G. Gaia~DR3 introduces an new data products, including the accurate radial velocity parameters for more than 33 million objects \citep{Katz2022}. The typical proper motion uncertainty respectively are 0.07\,\masyr for G$\approx$17\,mag, 0.5\,\masyr for G$=$20\,mag, the parallax uncertainty are 0.07\,mas at G$\approx$17\,mag, 0.5\,mas for G$=$20\,mag, and the mean G-band photometry uncertainty are 1\,mmag at G$\approx$17\,mag, 6\,mmag at G$=$20\,mag. The median formal precision of the velocities for the brightest, most stable Gaia stars lies at about 0.12\,\kms to 0.15\,\kms and smoothly increases for fainter stars \citep{Katz2022}.
 
Gaia~DR3 data is used in this study to identify ASCC~127 members through clustering algorithms and to estimate their kinematic properties and ages. These data underpin the division of ASCC~127 into subgroups and the investigation of their spatial and dynamical evolution.

\subsection{Dust Map}
\label{sec:Dust Map}
\citet{Edenhofer2024} utilized distance and extinction estimates for 54 million nearby stars derivede from Gaia BP/RP spectral data to conduct a parsec-scale three-dimensional interstellar dust extinction map, covering the spatial distribution within 1.25~kpc of the Sun. This map provides a detailed depiction of the internal structure of hundreds of molecular clouds in the solar neighborhood, making it a valuable tool for studying of star formation, Galactic structure and young stellar population.

The dust map is used to identify low-density cavities and high-density molecular clouds associated with the Cepheus Flare. These structures are critical for studying feedback processes and their role in triggering sequential star formation.

\subsection{CO Map}
\label{sec:CO}
The CO map used in this work is released from the work of \citet{Dame2001}, who combined data from 37 individual surveys to produce a comprehensive gas map of the entire Milky Way. These surveys were conducted using the CfA 1.2-meter telescope and a similar instrument located at Cerro Tololo in Chile. The surveys cover nearly all large local molecular clouds at higher latitudes. This map provides valuable insights into the large-scale structures of the molecular component of the Galaxy.

The CO map traces molecular gas distributions and kinematics, allowing us to identify filaments and cavities in the Cepheus Flare region. This is essential for linking feedback processes to observed ISM features.

\subsection{H~\texttt{I} Map}
\label{sec:HI}

The HI4PI dataset is the combined result of two high-sensitivity, high-resolution all-sky HI $21$\,cm surveys \citep{HI4PICollaboration2016}: the Effelsberg–Bonn H~\texttt{I} Survey (EBHIS) \citep{Kerp2011,Winkel2016} and the Galactic All-Sky Survey (GASS) \citep{McClure-Griffiths2009,Kalberla2010,Kalberla2015}. The former was conducted using the 100-meter Effelsberg telescope, while the latter employed the 64-meter Parkes telescope. This survey provides highly sensitive spectral data with superior velocity (frequency) and angular resolution, enabling precise measurements of the velocity and spatial distribution of H~\texttt{I} gas in the Milky Way. And the HI4PI survey data provides valuable insights into the analysis of dynamic characteristics of atomic gas.

The H~\texttt{I} map complements the CO data by providing insights into the diffuse atomic gas component of the ISM. This is particularly useful for studying large-scale structures, such as the Cepheus Flare Shell and Loop~III.

\subsection{PARSEC 1.2S isochrone model}
\label{sec:parsec_isochrone_model}

In this work, The PARSEC (the PAdova \& TRieste Stellar Evolution Code) isochrones are retrieved from the CMD web interface v3.7\footnote{\url{http://stev.oapd.inaf.it/cgi-bin/cmd}}, and we choose a set of PARSEC isochrone model (release v1.2s) with metallicity $\text{[M/H]}=0.0$, and the isochrones selected range from $\tau/\text{Myr}=1$ to 60 with an interval of $\Delta\,{\tau/\text{Myr}}=1$.

Isochrones are fundamental tools for determining key parameters of stellar clusters, including their age, by fitting them to observed photometric data. However，it has been observed in prior researches that the low-mass segment of the PARSEC 1.2S isochrones does not align well with the photometric observations of stellar groups when plotted on color-magnitude diagrams(CMDs) \citep[see e.g., ][]{Castellani2001, Bell2015MN, Li2020ApJ}. To address this issue, \citet{Wang2025} conducted a detailed analysis using three benchmark stellar clusters(Hyades, Pleiades, and Praesepe)to quantify the differences between observational data and theoretical predictions. Their work led to the development of empirical color correction functions specifically tailored for the PARSEC 1.2S isochrone models. These color-corrected isochrones models will be used in this work.

\section{Membership}
\label{sec:membership}

In this section, we describe the identification, refinement, and grouping of ASCC~127 members, as well as the estimation of their ages and masses. The sky region and sample sources of the member star identification are based on the prior work of \citet{Kounkel(2019)} (abbreviated as KC19). We re-identified candidate members of the moving group ASCC~127 using the Friends-of-Friends (FOF) clustering algorithm, and then applied the K-means method to divide these members into the subgroups.

\begin{figure}
    \centering
        {%
            \includegraphics[width=0.95\linewidth,trim=0.1cm 0.4cm 0cm 2.0cm, clip]{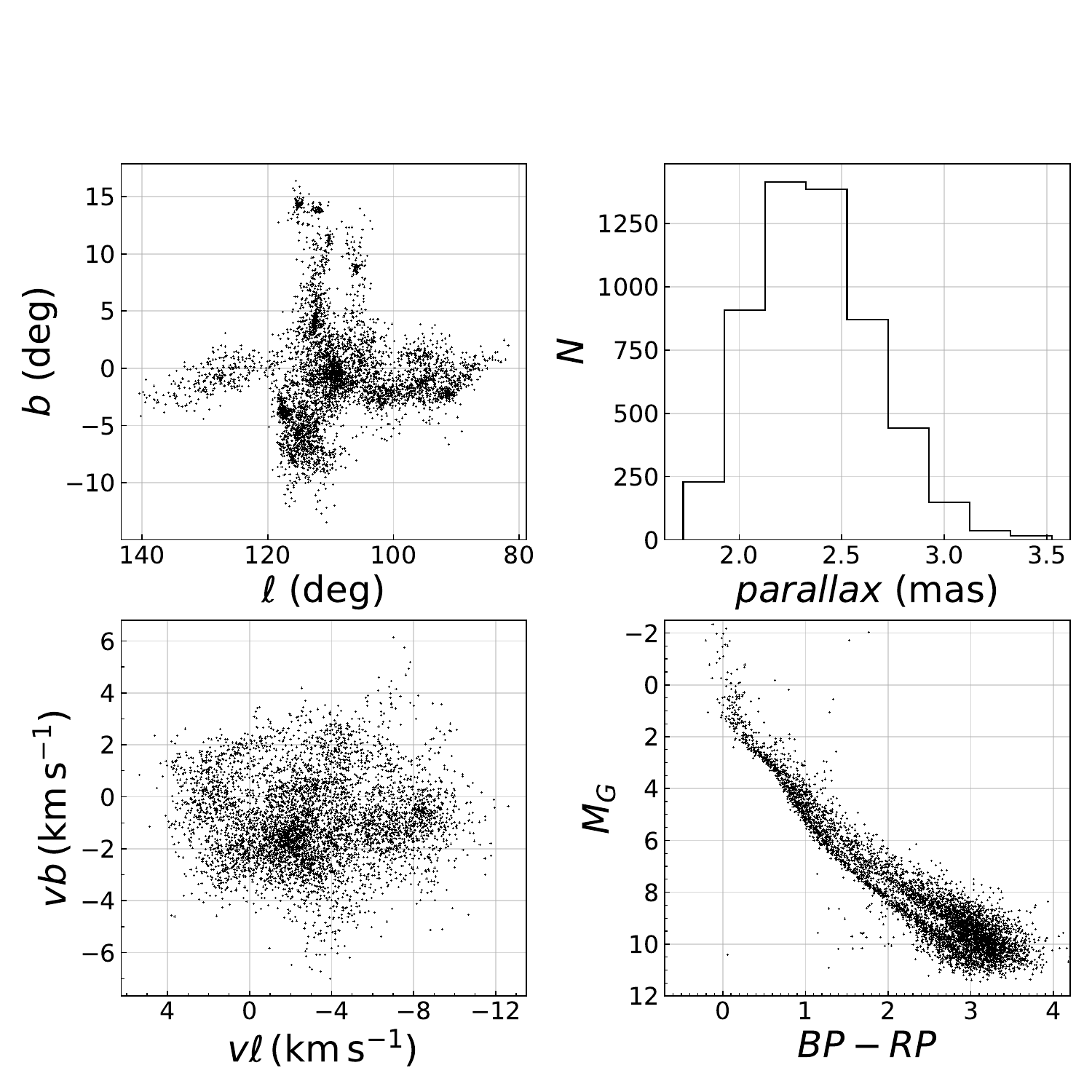}
            }

\caption{Distributions of 5,450 member candidates (black dots), identified by the Friends-of-Friends (FoF) algorithm, in five-parameter space (Galactocentric coordinate ($\ell$, b), parallax, tangential velocity (v$\ell$, vb)) and the CMD ($M_G$ vs. BP$-$RP).
\label{fig:fof_first}}
\end{figure}

\begin{figure}  
    \centering
        {%
            \includegraphics[width=0.45\linewidth, trim=0.3cm 0.2cm 0.2cm 1.5cm, clip]{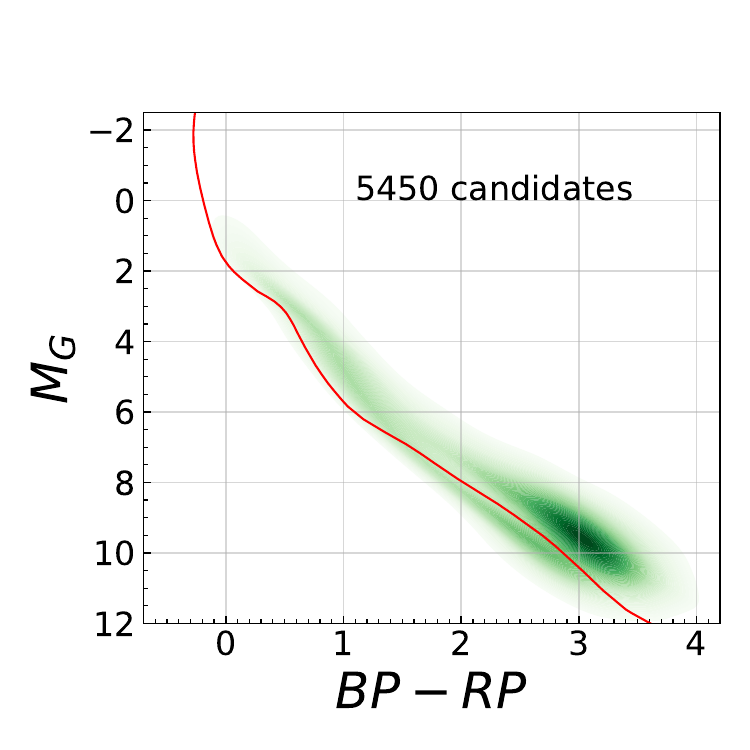}\quad
            \includegraphics[width=0.45\linewidth, trim=0.3cm 0.2cm 0.2cm 1.5cm, clip]{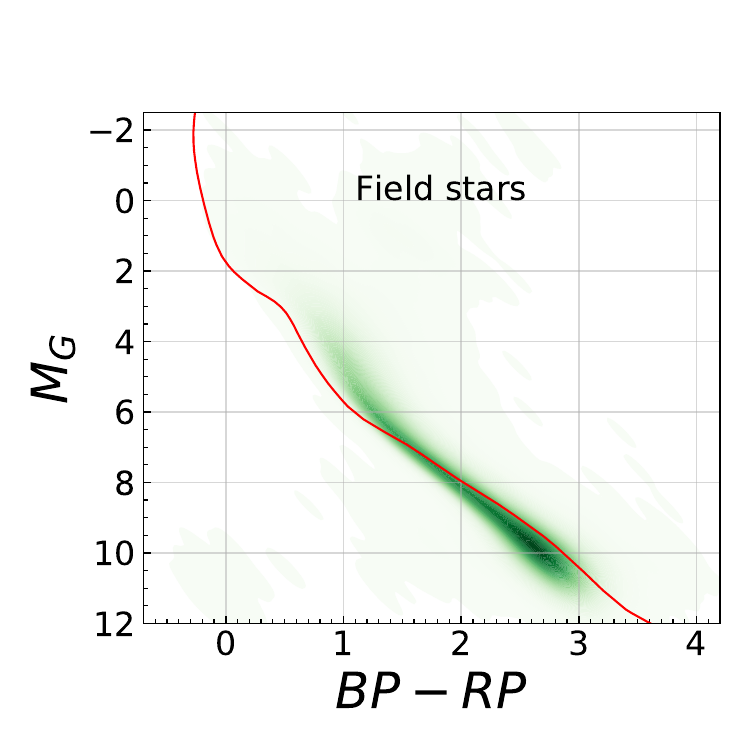}}

\caption{Comparison the 5,450 member candidates with the field stars in this region. The left subplot: the density map of the 5,450 member candidates. The right subplot: the density map of the field stars in this region. The red dashed line is the corrected-color isochrone from PARSEC model with an age of 50~Myr and solar metallicity ($[M/H]$=0).
\label{fig:5822_vs_bg}}
\end{figure} 

\begin{figure}
    \centering
        {%
            \includegraphics[width=0.95\linewidth, trim=0.2cm 0.2cm 0.2cm 1.2cm, clip]{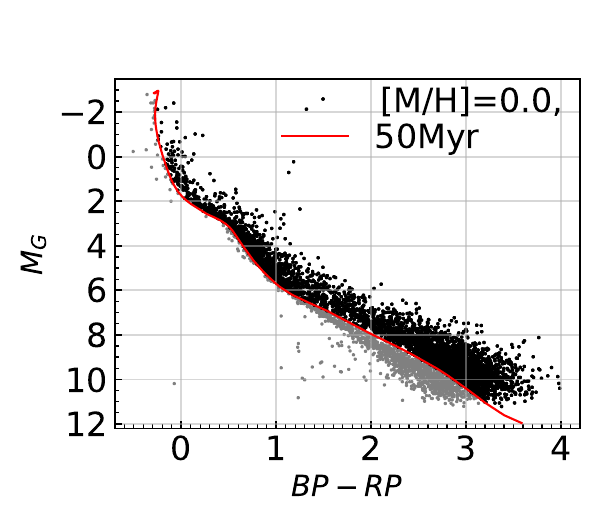}
            }

\caption{Comparison the results before and after noise elimination in CMD. The black dots are the 4,140 remaining member candidates, while the grey dots indicate the 1,310 field stars that were removed.
\label{fig:cmd_cut}}
\end{figure}

\begin{figure}
    \centering
        {%
            \includegraphics[width=0.5\linewidth,trim=0.2cm 0.2cm 0.2cm 1.4cm, clip]{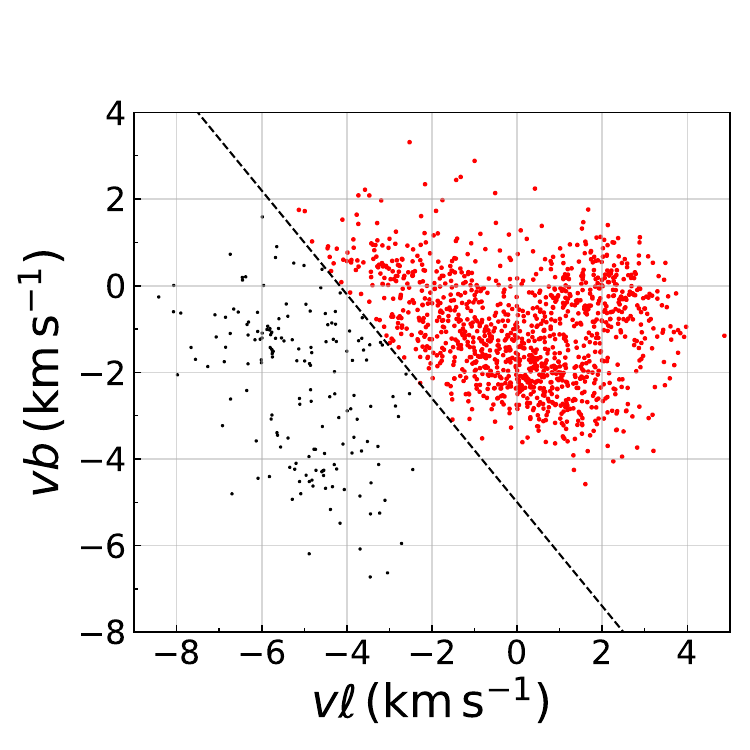}\quad
            \includegraphics[width=0.4\linewidth,trim=0.2cm 0.2cm 0.2cm 1.4cm, clip]{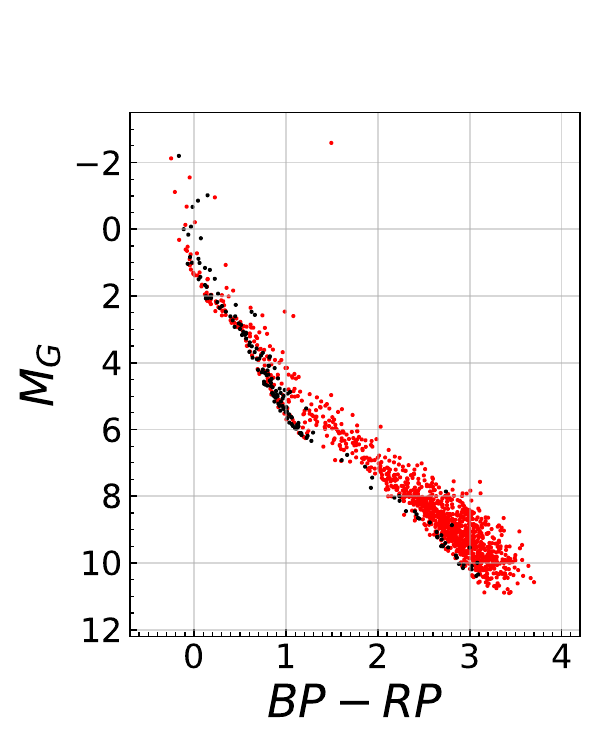}
            }
\caption{Comparison of tangential velocity distribution for Group~3 before and after noise elimination. Group~3 is divided into member candidates (red dots) and noise (black dots).
\label{fig:G3_vlb_cut}}
\end{figure}

\setcounter{table}{0}

\setcounter{table}{0}
\begin{table*}
\scriptsize
\begin{center}
\caption{Members of Groups\,1$-$5. The full table of all identified members in Groups~1$-$5 is available in a machine-readable format.}
\label{members of Gs5} 

    \begin{tabular}{cccccccccc}
        \hline 
        \hline
        \textbf{$\ell$} & \textbf{b} & \textbf{plx} & \textbf{${\mu_{l^*}}$} & \textbf{${\mu_b}$} & \textbf{BPmag} & \textbf{RPmag} & \textbf{Gmag} & \textbf{$A_{V}$} & \textbf{Group} \\ 
        (degree)& (degree)& (mas) &(\masyr)& (\masyr)& (mag)& (mag) &(mag)&(mag) & \\     
        \hline
        126.58688&	1.04810&	2.55&	0.07&	0.47&	12.51&	11.53&	12.11&	0.43&	1 \\
        126.55740&	1.08007&	2.63&	-0.50&	1.65&	15.51&	13.66&	14.61&	0.38& 	1 \\
        127.39135&	1.99943&	2.59&	0.06&	0.63&	12.51&  11.58&	12.13&	0.34&	1  \\
        132.19980& 2.92269&	2.51&	0.92&	0.00&	10.14&	9.91&	10.06&	0.56&	1  \\
        135.87525&	6.06720&	3.19&	1.58&	1.36&	10.34&  9.70&	10.10&	0.26&	1  \\ 
        \hline
    \end{tabular} 
\end{center}       
\footnotesize{The Galactic coordinates (Column\,1$-$2), parallaxes (Column\,3) of the members in each group, as well their proper motions (Column\,4$-$5) in the Galactic coordinates with respect to LSR, observed Gaia photometric data (Column\,6$-$8), V$-$band extinction (Column\,9) and group IDs (Column\,10). } 
\end{table*}

\begin{figure*}
    \centering
        {%
            \includegraphics[width=0.95\linewidth, trim=0.0cm 0.3cm 0.cm 1.8cm, clip]{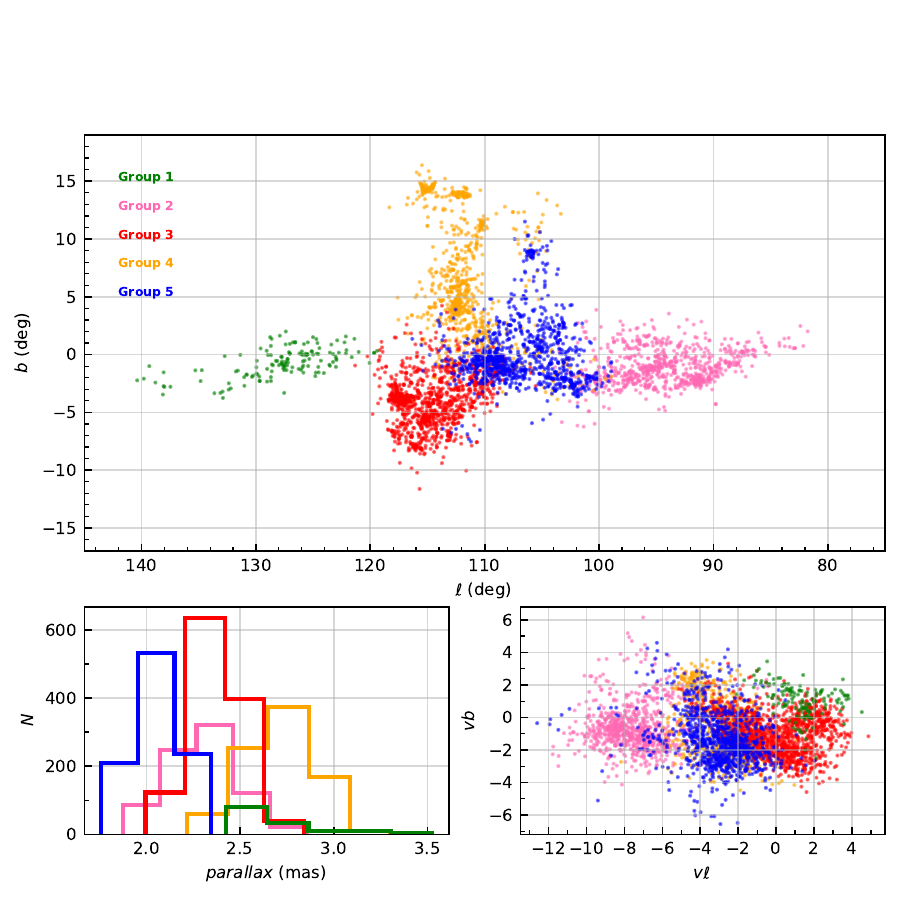}\quad
            }
\caption{Distribution of Groups\,1$-$5 in the Galactocentric coordinate ($\ell$, b), parallax, tangential velocity (v$\ell$, vb). In all of the panels, green dots, hotpink dots, red dots, orange dots and blue dots  represent the candidate members of Group\,1, Group\,2, Group\,3, Group\,4 and Group\,5, respectively.
\label{fig:5grps_memprobs}}
\end{figure*}

\subsection{Target Selection}
\label{sec:Target Selection}

To re-identify the membership of ASCC~127, we utilize high-precision astrometric and photometric data from Gaia~DR3. We select a sample using the following criteria:

(1) The Galactic longitude ($\ell$) ranging from \(70^\circ\) to \(150^\circ\), the Galactic latitude (b) ranging from \(-30^\circ\) to \(25^\circ\).
    
(2) For all sources within our 100$-$800~pc selection, we employed the geometric distance \( d \) from \citet{Bailer_Jones2021}, with the associated uncertainty \( \sigma_d \)\footnote{\( \sigma_d \) is only used for X$-$Y trajectory tracing in Section~\ref{sec:relationship each group}.} calculated as the average of the lower and upper 1\( \sigma \) deviations\footnote{Note that unless otherwise specified, these definitions hold throughout this work.}. To ensure high-quality parallax measurements, only sources with a parallax uncertainty \(\sigma_\varpi < 0.5\)~mas and \(\varpi / \sigma_\varpi  > 10\) were selected.
    
(3) The proper motion angular velocities \((\mu_{\alpha*}, \mu_\delta)\) (in \(\mathrm{mas\,yr^{-1}}\)) were converted to tangential velocities \((v_{\alpha*}, v_\delta)\) \footnote{
\(
v_{\alpha*}/\mathrm{km\,s^{-1}} = 4.74 \times \mu_{\alpha*}/\mathrm{mas\,yr^{-1}} \times d/\mathrm{pc}, 
\)
\(
v_\delta/\mathrm{km\,s^{-1}} = 4.74 \times \mu_\delta/\mathrm{mas\,yr^{-1}} \times d/\mathrm{pc}
\)
} (in \(\mathrm{km\,s^{-1}}\)).
The selected ranges for tangential velocities were \(-5 < v_{\alpha*}/\mathrm{km\,s^{-1}} < 25\) and \(-25 < v_\delta/\mathrm{km\,s^{-1}} < 12\).

(4) To ensure data quality, we excluded sources with
$\mathrm{astrometric\_excess\_noise} \geq 1$,
which filters out non-single stars and low-quality astrometric measurements.

(5) To ensure the selected stars with high quality in astrometric parameters (position, parallax and proper motion), we required the maximum standard deviation of the 5$-$parameter solution (\(\mathrm{astrometric\_sigma5d\_max}\)) to be less than $0.3$.

(6) To limit to sources with high quality astrometric solutions, we required the Renormalized Unit Weight Error (RUWE) to be less than $1.4$.

(7) To mitigate parameter uncertainties caused by insufficient observations, as more visibility periods reduce random errors and improve solution accuracy, only sources with more than 8 visibility periods (\(\mathrm{visibility\_periods\_used}\)) were included.   

(8) To maintain high-quality photometry, we required the uncertainty in the Gaia \(G\)$-$band magnitude, $\sigma_{G}$, to be less than 0.03~mag.

(9) The stellar color \(BP-RP\) was restricted to the range \(-0.6\) to $4.3$, and the absolute magnitude \(M_G\) was required to lie between \(-4\) and 13~mag, to exclude anomalous data and non-stellar sources, 

Criteria (1)$-$(3) and (9) adopt the known properties of ASCC 127 from KC19, while criteria (4)$-$(8) follow Gaia quality control conditions specified in KC19. Employing the above criteria, we obtain 977,337 stars for the further member refinement of ASCC~127.


\subsection{Membership refinement}
\label{sec:Mem}

We adopt the FOF algorithm using the software {\tt ROCKSTAR} \citep{Behroozi(2013)} to search for members of the ASCC\,127. {\tt ROCKSTAR} employs a technique of adaptive hierarchical refinement in 6D phase space. It divides all the stars into several FOF groups by tracking the high number density clusters and excising stars that are not grouped in the star aggregates. Using the 5D phase information (i.e., $\ell$, b, $\mu_{\ell^*}$\footnote{The proper motion component in Galactic longitude, $\ell$, with the local standrd of rest (LSR).}, $\mu_b$\footnote{The proper motion component in Galactic latitude, b, with LSR.}, and $d$) of each star as the input parameters (noting that radial velocity (RV) is set to zero for each star), the optimized {\tt ROCKSTAR} \citep{TianH_PHD} will automatically adjust the linking-space between members of "friend" stars, and divide them into several groups, simultaneously removing isolated individual stars from the groups.

In this step, we obtained 5,450 candidate members, as shown in Figure~\ref{fig:fof_first}. From the CMD subplot of this figure, we can observe two distinct branches, one for the pre-main sequence stars and the other for the field stars. To enhance the distinction between the two populations, we display the Hess diagram for the CMD of ASCC~127, and all the analyzed stars (mostly field stars) in Figure~\ref{fig:5822_vs_bg}. It is obvious that the candidate members of ASCC~127 are contaminated by the field stars and a 50~Myr isochrone from the PARSEC model \citep{Bressan2012, Chen(2019)} with empirical color correction with the solar metallicity seems to separate the two branches in the CMD very well.

\begin{figure*}
    \centering
        {%
            \includegraphics[width=0.188\linewidth, trim=0.1cm 0.2cm 0.35cm 1cm, clip]{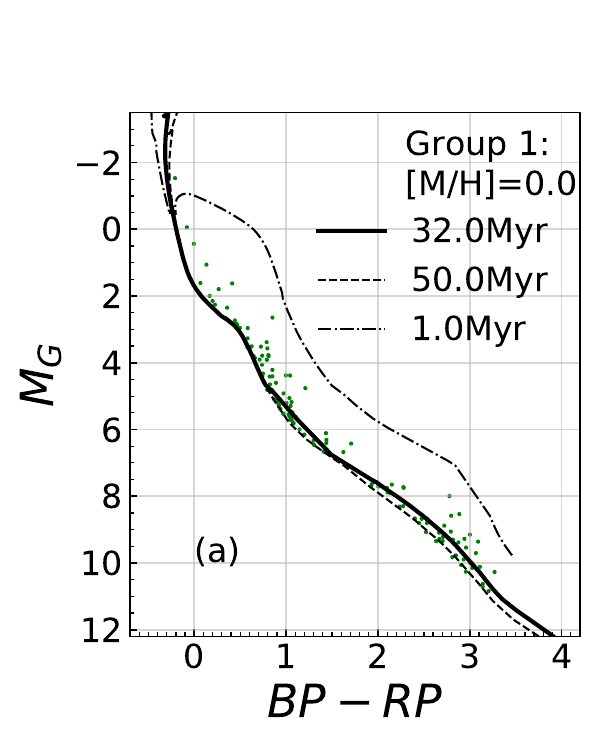} 
            \includegraphics[width=0.147\linewidth, trim=2.2cm 0.2cm 0.35cm 1cm, clip]{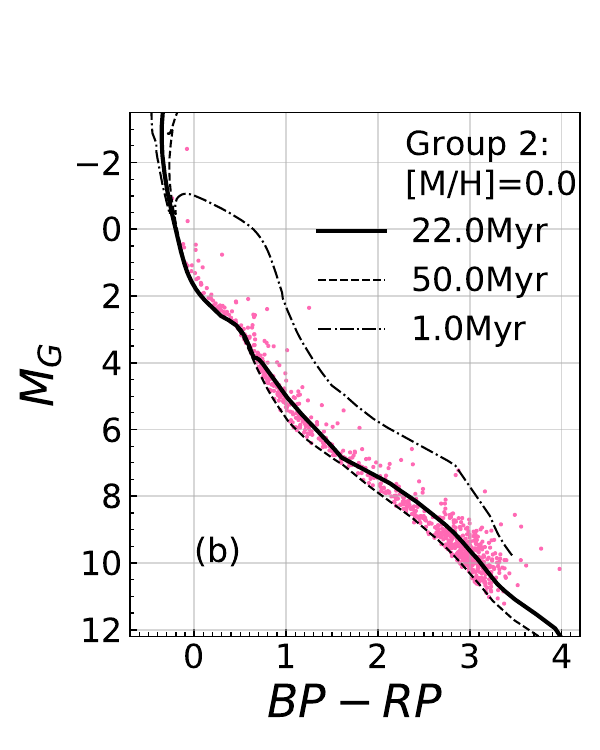}
            \includegraphics[width=0.147\linewidth, trim=2.2cm 0.2cm 0.35cm 1cm, clip]{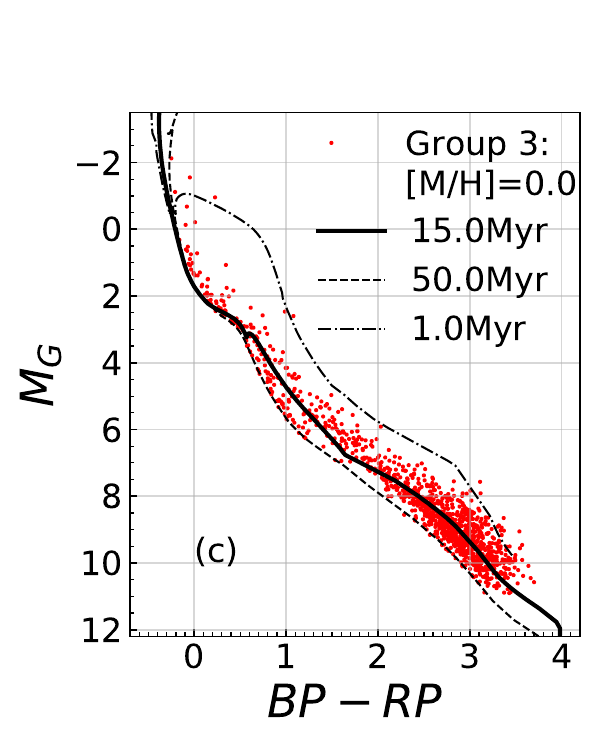} 
            \includegraphics[width=0.147\linewidth, trim=2.2cm 0.2cm 0.35cm 1cm, clip]{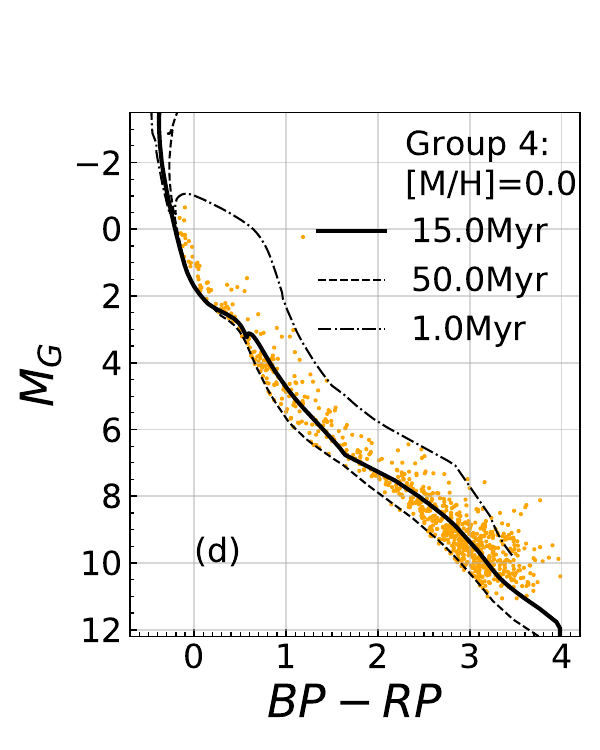}       
            \includegraphics[width=0.147\linewidth, trim=2.2cm 0.2cm 0.35cm 1cm, clip]{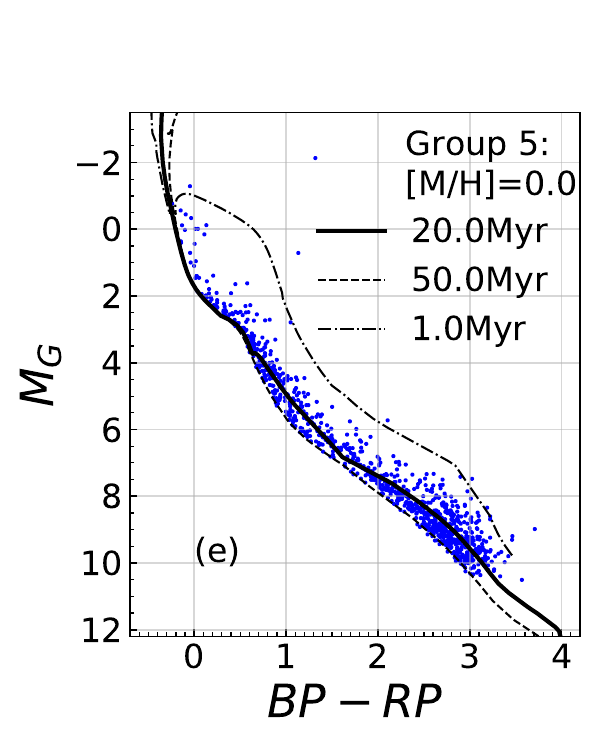} 
            }  
            
\caption{CMDs for Groups\,1$-$5 (same color-coding as in Figure \ref{fig:5grps_memprobs}). The black solid, dotted-dashed and dashed curves in each panel represent the best-fit, 50~Myr and 1~Myr PARSEC isochrones with empirical color correction, respectively. 
\label{fig:5grps_CMD_bestfits}}
\end{figure*}

The CMDs shown in Figures~\ref{fig:fof_first} and \ref{fig:5822_vs_bg} are not dereddened. To better separate the candidate members and the field stars, we dereddened the Gaia photometry of the 5,450 candidates using the the three-dimensional extinction map STructuring by Inversion the Local Interstellar Medium (STILISM\footnote{https://stilism.obspm.fr}; \citet{Lallement2014AA91L, Capitanio2017AA65C, Lallement2018AA132L}), similar as done in \cite{Wang2025}. From the STILISM, we obtain the  color excess $E(B-V)$  for each candidate and then calculate the visual extinction as \( A_{V} = 3.1 \times E(B-V) \). We construct a grid of Gaia BP$-$RP colors, extinction in BP, RP, and G bands ($A_{BP}$, $A_{RP}$, $A_G$) for main squence stars with spectral type ranging from O5$-$M6 with $A_{V}$ ranging from 0 to 10~mag.  Employing the grid, we obtain the $A_{BP}$, $A_{RP}$, $A_G$ for each source using its  observed BP$-$RP color and the $A_{V}$ from the STILISM. The Gaia photometry of each source is derendened using these extinctions.
 
In Figure~\ref{fig:cmd_cut}, we show the deredened CMDs for the candicate members of ASCC~127. We decontaminate the field stars using the above  50\,Myr PARSEC isochrone of solar metallicity with empirical color correction. In this way, we select 4,140 (75\%) candidate members located above this isochrone for the further study. We must stress that our approach to the selection is rather rough. There are still some field stars in the sample, which are further excluded in Section~\ref{sec:Groups} based on the CMDs and the proper motions.

\begin{figure*}
    \centering
        {%
            \includegraphics[width=0.90\linewidth, trim=0.1cm 0.1cm 0.1cm 0.1cm, clip]{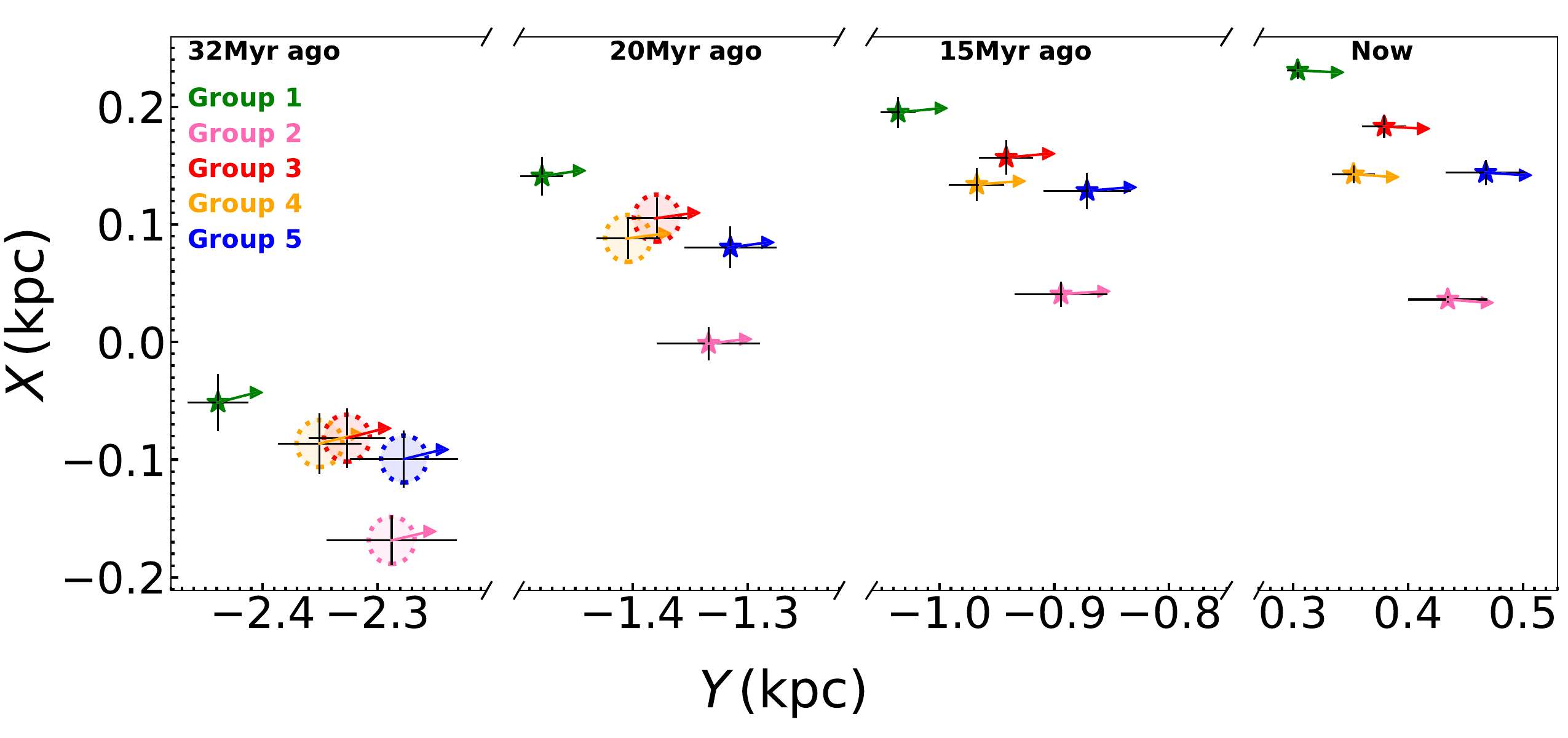} 
            }  
            
\caption{2D trajectory plot ($X$, $Y$) of the backwards orbit integrations of Groups~1$-$5, tracing their motion from the past (32~Myr ago) to the present in Heliocentric Cartesian coordinates ($X$, $Y$, $Z$). The open circle with a dotted outline represents that the group had not yet formed, while the star denotes that the group had formed. These symbols are placed at specific times: 32, 20, and 15~Myr ago, and at present. The errorbar represent the positional dispersion in ($X$, $Y$), and the arrows indicate the median motion direction of each group.
\label{fig:traceback_Myrs}}
\end{figure*}

\subsection{Grouping}
\label{sec:Groups}

We employed the K-means clustering method to divide the ASCC~127 into subgroups. We start to group the 764 sources with the RV uncertainties $\leq$5\,\kms. The grouping are performed using the three-dimensional Cartesian coordinates $(x, y, z)$ and the velocity in their direction $(U, V, W)$ with LSR. Using the K-means clustering method, these sources can be divided into 5 groups. For these sources without RVs, we divide them into individual groups using the five-dimensional spaces, including the three-dimensional Cartesian coordinates and proper motions. A source is assigned to a group if the distance in the five-dimensional spaces from the source to the center of the group is minimized.

We notice that Group~3 could be still contaminated by the field stars. In Figure~\ref{fig:G3_vlb_cut}, we show the distribution of the tangential velocity  (v$\ell$ , vb)\footnote{\(v_{\ell*}/\mathrm{km\,s^{-1}} = 4.74 \times \mu_{\ell*}/\mathrm{mas\,yr^{-1}} \times d/\mathrm{pc} \), \(v_{b}/\mathrm{km\,s^{-1}} = 4.74 \times \mu_{b}/\mathrm{mas\,yr^{-1}} \times d/\mathrm{pc} \)}  in the Galactic coordinate and the CMD of Group~3. There are a group of stars with the tangential velocities which is deviated from the majority of the sources in Group~3. In the left panel of Figure~\ref{fig:G3_vlb_cut}, we simply remove these 169 sources based on the distribution of their tangential velocities. In the CMD (right panel of Figure~\ref{fig:G3_vlb_cut}), these sources are located at the bottom of the locus where the Group~3 sources are located, and likely to be contaminated by field stars. In comparison, \citet{Pang2022} refined the membership of ASCC~127, focusing particularly on Group~3 in our study, and identified that 49\% of the sources in Group~3 belong to the refined membership. For the other groups, we have not noticed evident contamination in the members. Finally, we obtained a total of 3,971 members ($\sim$73\% of the initial 5,450 candidates), with the member candidates distributed across Groups\,1$-$5 as follows: 140, 796, 1194, 864 and 977, respectively\footnote{The complete list of all members of Groups~1$-$5 identified in this work is available in a machine-readable form.}. The total number of members increases by a factor of 2 than the previous ones in KC19. A detailed comparison of the membership in this work and in KC19 is presented in Appendix~\ref{Appen:com}.

In Figure~\ref{fig:5grps_memprobs}, we show the spatial distribution of the five groups, as well the distributions of their parallax and tangential velocities in Galactic coordinate. In the top of Figure~\ref{fig:5grps_memprobs}, Groups~1$-$5 exhibit a loose and extended morphology in the two-dimensional projection of the Galactic coordinate. In the lower left of Figure~\ref{fig:5grps_memprobs} shows that the parallax distribution peaks for Groups~1$-$5 are similar, all falling between 2.0 and 2.7~mas. In the bottom right of Figure~\ref{fig:5grps_memprobs}, we can observe that the distribution of tangential velocities among stars in Groups~1$-$5 exhibits distinct patterns. Apart from Group~1, most members in Groups~2$-$5 share a similar trend with their tangential velocity components v$\ell$ and vb generally being negative. This indicates that these members tend to move towards lower Galactic latitudes. In contrast, nearly all member stars in Group~1 exhibit positive v$\ell$ and vb values, suggesting that members in this group primarily move towards higher Galactic longitudes and latitudes. And there are slightly differences in these groups, which will be discussed in Section \ref{sec:relationship each group}. The median values of the positions, distances, and tangential velocities of Groups~1$-$5 are summarized in Table~\ref{properties of Gs5}.

\subsection{Ages and Masses} 
\label{sec:age}

Isochrone fitting is a typical method to estimate stellar ages. To obtain the precise age, we choose a set of PARSEC isochrone model (release v1.2s), which is described in detail in Section~\ref{sec:parsec_isochrone_model}. As described in Section~\ref{sec:parsec_isochrone_model}, there is a discrepancy between the low-mass end of the PARSEC isochrones and the observed photometric data of stellar groups in CMDs. This discrepancy could lead to an underestimation of the derived age, particularly for young stellar groups. Therefore, to obtain a more accurate age for stellar group, we apply the empirical color correction to the PARSEC isochrones, as proposed in \citet{Wang2025}. By using these isochrones with empirical color corrections and employing the method described in \citet{Liu(2019)}, we obtain the age of each group by minimizing the mean distance of the group member between their locations and the isochrones in the $M_{G}$ vs. BP$-$RP CMDs.

In Figure~\ref{fig:5grps_CMD_bestfits}, we display the CMDs for Groups~1$-$5. The black solid curve in each panel is the best fit isochrone for each group. Group~1, that is 32$^{+4}_{-5}$~Myr, is slightly older than the other groups. Groups~2 and 5 are younger, their ages being 22$^{+4}_{-1}$ and 20$^{+2}_{-1}$~Myr, respectively. Groups~3 and 4 are the youngest and have a same age (15$^{+1}_{-2}$~Myr). The uncertainty in age is estimated by performing bootstrap resampling on the dataset of BP$-$RP colors and absolute magnitudes $M_G$, where each resampled dataset consists of half of the original data. These results have been listed in Table~\ref{properties of Gs5}\footnote{For comparison, using the uncorrected PARSEC 1.2S isochrones and employing the method described above, we obtained the ages of Groups~1$-$5: 26$^{+5}_{-2}$, 17$^{+1}_{-1}$, 10$^{+1}_{-1}$, 10$^{+1}_{-1}$, 14$^{+1}_{-1}$~Myr, respectively. The empirical color correction applied to the PARSEC 1.2S isochrones\citep{Wang2025} shifts them to redder colors in this color range of BP$-$RP greater than 0.3~mag}, resulting in slightly older age estimates compared to the uncorrected ones. However, the relative age sequence among Groups~1$-$5 remains unchanged, with Group~1 being the oldest, Groups~2 and 5 relatively older, and Groups~3 and 4 the youngest.

We calculate the current mass of each member in Groups\,1$-$5 by comparing its location with the PARSEC isochrones after an empirical color correction on the CMDs. The total masses of the members for Groups~1$-$5 are approximately $\sim$140~\Msun, $\sim$600~\Msun, $\sim$700~\Msun, $\sim$560~\Msun, $\sim$800~\Msun, respectively. These results are also listed in Table~\ref{properties of Gs5}.


\begin{figure*}[!ht]
    \centering   
    \includegraphics[width=1.8\columnwidth, trim=0.2cm 0.0cm 0.4cm 1.8cm, clip]{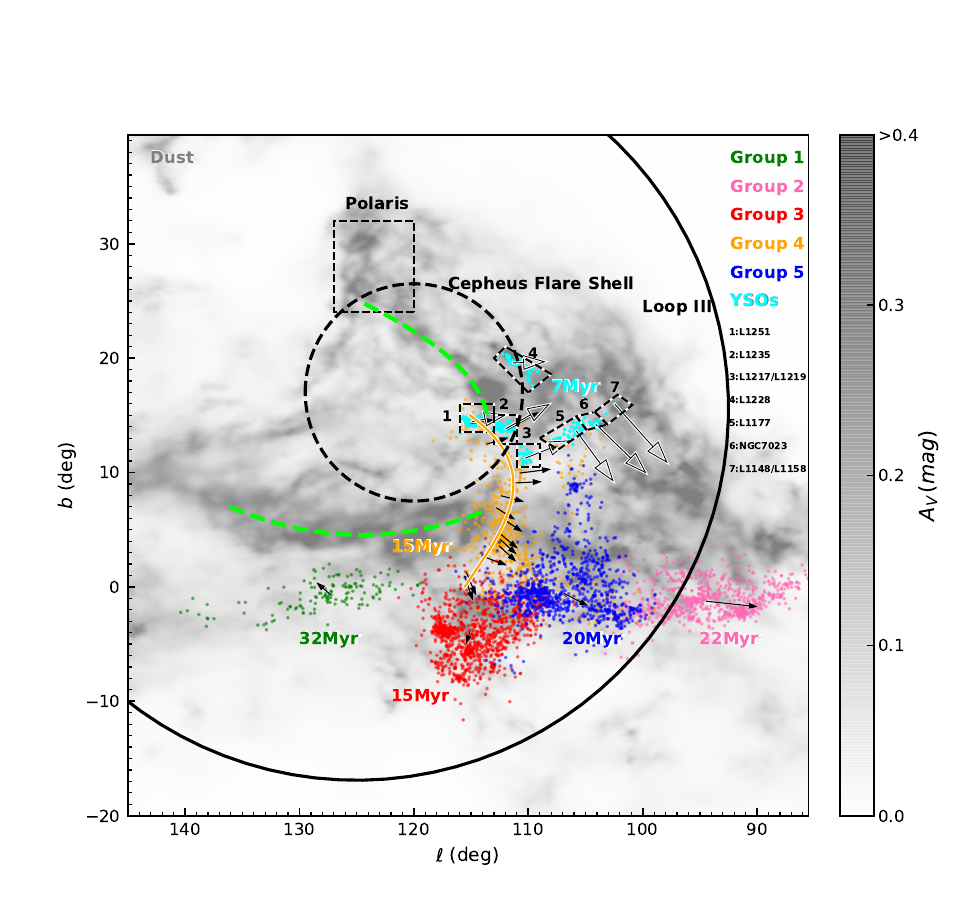}
    \caption{Dust extinction map from \citet{Edenhofer2024}, displayed in Galactic coordinates. The black dashed circle marks the Cepheus Flare Shell, provided by \citet{Kun2009}, and the black solid circle indicates Loop~III, provided by \citet{Kun2007}. The black dashed rectangle outlines the nearby star-forming region Polaris. The lime dashed line marks cavity, and cyan points represent YSOs from \citet{Szilagyi2021MN}, 
    with black dashed lines enclosing YSOs associated with seven different molecular clouds. Stellar groups are marked in different colors (as shown in Figures~\ref{fig:5grps_memprobs} and \ref{fig:5grps_CMD_bestfits}). The distribution trend of Group~4 is outlined with yellow color and thick white solid lines. Black arrow shows the direction of median proper motion of group members, with the length indicating its value. 
     \label{fig:Gs1-5-ext}}
\end{figure*} 

\section{Discussion}
\label{sec:discu}

ASCC~127 is located at the similar distance to the Cepheus Flare star-forming region. Compared with the Cepheus Flare, ASCC~127 is relatively older. Exploring the relationship between ASCC~127 and the Cepheus Flare will enhance our understanding of star formation, as well as the interaction between feedback from the "older" ASCC~127 population and the ISM in this region. Here below we discuss the relationships among the ASCC~127 subgroups, their connection to the Cepheus Flare star-forming region, and the role of feedback mechanisms in shaping the ISM and triggering sequential star formation.

\begin{figure}
    \centering
        {%
            \includegraphics[width=0.90\linewidth, trim=0.cm 0.2cm 0.30cm 1.8cm, clip]{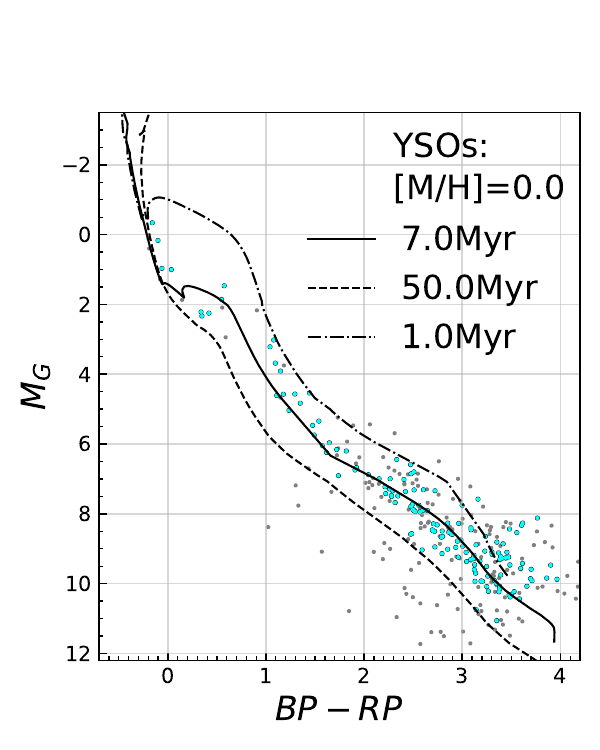} 
            }  
            
\caption{CMDs for 319 nearby young stellar objects (YSOs, grey and cyan dots), including 149 YSOs (cyan dots) that are part of Group~4. The black solid, dotted-dashed and dashed curves in each panel represent the best-fit, 50~Myr and 1~Myr PARSEC isochrones with empirical color correction, respectively. 
\label{fig:yso_CMD_bestfits}}
\end{figure}

\subsection{Relationship among the groups}
\label{sec:relationship each group}

To understand the spatial and temporal relationships among the ASCC~127 subgroups, we reconstruct the past orbits of groups using the  gravitational potential model of the Milky Way (MWPotential2014), as implemented in the galpy Python package developed by \citet{Bovy(2015)}. This analysis provides insights into the group dynamics and their role in sequential star formation.

Using the median values of six-dimensional parameters ($\ell$, $b$, $\mu_{\ell^*}$, $\mu_b$, $d$, and RV$_{LSR}$\footnote{RV$_{LSR}$ is radial velocity (RV) with respect to the local standard of rest (LSR).}) for Groups~1$-$5, we integrate their past orbits in the Heliocentric Cartesian coordinate system $(X, Y, Z)$ from the present day ($t=0$) to 32~Myr ago, with a time step of 1~Myr. 
To estimate the uncertainty in the orbit integrations, we further consider the standard deviations of these six parameters ($\sigma_{\ell}$, $\sigma_{b}$, $\sigma_{\mu_{\ell^*}}$, $\sigma_{\mu_b}$, $\sigma_{d}$, and $\sigma_{RV_{LSR}}$). Since the uncertainties in $\ell$ and $b$ ($\sigma_{\ell}$ and $\sigma_{b}$) have a negligible impact on the orbit integrations, we neglect them in our analysis. Thus, we only account for the effects of the standard deviations of $\mu_{\ell^*}$, $\mu_b$, $d$, and RV$_{LSR}$ ($\sigma_{\mu_{\ell^*}}$, $\sigma_{\mu_b}$, $\sigma_{d}$, and $\sigma_{RV_{LSR}}$), assuming that these uncertainties follow independent Gaussian distributions. Based on this assumption, we generate 10,000 realizations of these parameters by randomly sampling $\mu_{\ell^*}$, $\mu_b$, $d$, and RV$_{LSR}$ from their respective Gaussian distributions. For each realization, we integrate an orbit, resulting in a total of 10,000 integrated orbits. We then determine the median trajectory from these sampled orbits and use the 1$\sigma$ standard deviation to quantify the uncertainty range of the integrated orbits. We then derive the median orbit from the sampled trajectories and use the 1$\sigma$ standard deviation to quantify the uncertainty range of the integrated orbits.
For this integration, we adopt a solar motion of $(U, V, W) = (11.1, 12.24, 7.25)\,\text{km}\,\text{s}^{-1}$ \citep{Schonrich2010} and a solar position of $(X, Y, Z) = (0, 0, 5.5)\,\text{pc}$ \citep{Reid2019}. In this coordinate system, $X$ increases toward the Galactic anticenter, and $Y$ follows the direction of Galactic rotation. We also estimate the trajectories of the parent molecular clouds under the assumption that their initial kinematics were similar to those of the young groups at the time of formation. 

Figure~\ref{fig:traceback_Myrs} presents the backward orbital integrations in the ($X$, $Y$) plane for Groups~1$-$5. The $Z$-direction is not shown due to the absence of clear variations among the groups. The figure also marks the estimated formation status of each group at 32, 20, and 15~Myr ago, based on the best-fit ages derived in Section~\ref{sec:age}. Different symbols (e.g., open circles and stars) are used to distinguish formed and unformed groups, while errorbars represent the positional dispersions and arrows indicate their motion directions. The traceback analysis reveals that the relative spatial distribution of Groups~1$-$5 has evolved over time. A more detailed discussion of these changes follows.

Approximately 32~Myr ago, the oldest group, Group~1 (32~Myr), formed, while Groups~2$-$5 had not yet emerged. At that time, Group~1 was located $\sim$110~pc nearer to the parent molecular clouds of  Groups~3 and 4, compared to the approximately 180~pc separating it from the parent clouds of Groups~2 and 5. The distances among the parent molecular clouds for Groups~2$-$5 are from 60 to 100~pc. 

Groups~2 and 5 formed approximately 17~Myr after Group~1. Initially, the distance between these two groups was about 75~pc at the time of their formation; today, this distance has increased to around 106~pc.  

The two youngest groups, Groups~3 and 4, formed approximately 15 million years ago, which is after the formation of Groups~2 and 5. Since their formation, Groups~3 and 4 have been co-moving while maintaining a constant distance of about 70~pc between them.

Generally, Groups~1$-$5 are formed within approximately 100~pc of each other, whereas Group~1 is located at a greater distance from the other groups.

\begin{figure*}[!t]
    \centering
    \includegraphics[width=1.8\columnwidth, trim=0.2cm 0.0cm 0.4cm 1.8cm, clip]{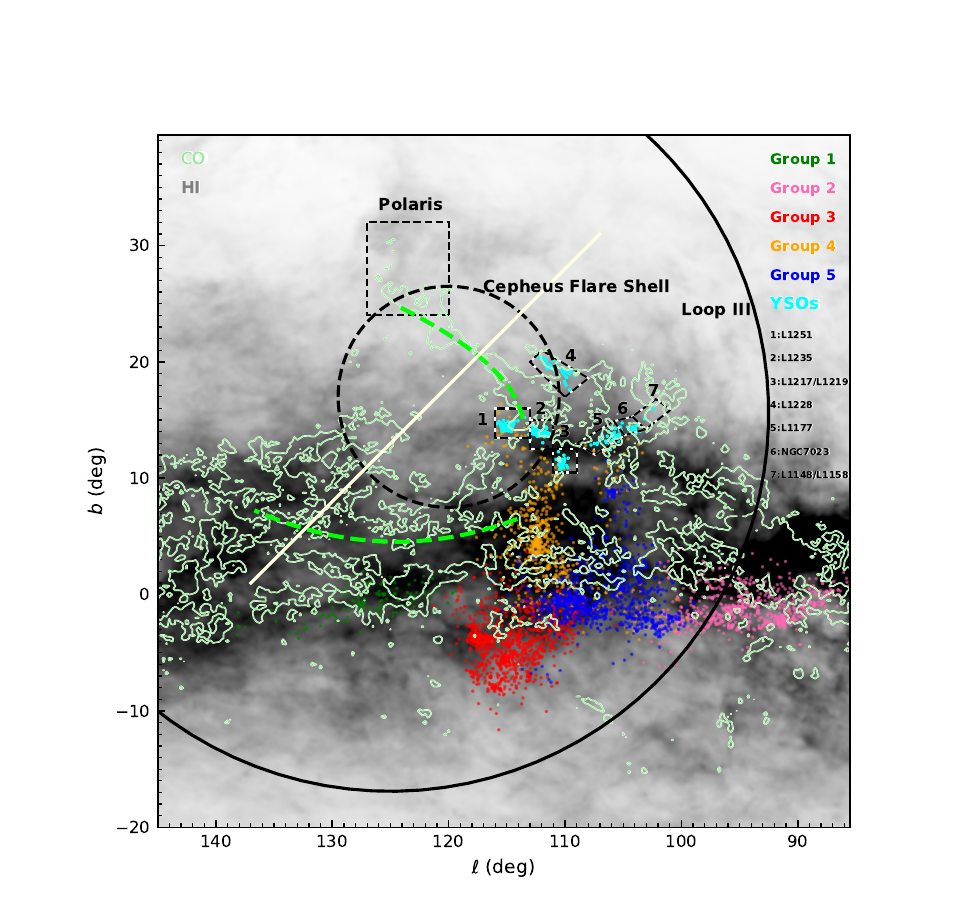}

    \caption{Left panel: CO integrated intensity map (lime contour) from \citet{Dame2001}, and HI integrated intensity map (gray background) from \citet{HI4PICollaboration2016}. The light-yellow line with the direction from upper-right to lower-left shows the routing of the position-velocity diagram of the cavity. 
    \label{fig:Gs1-5-CO-HI}
    }
\end{figure*} 
\begin{figure*}[!t]
    \centering
    \includegraphics[width=0.48\linewidth, trim=.2cm 0.0cm 0.5cm .2cm, clip]{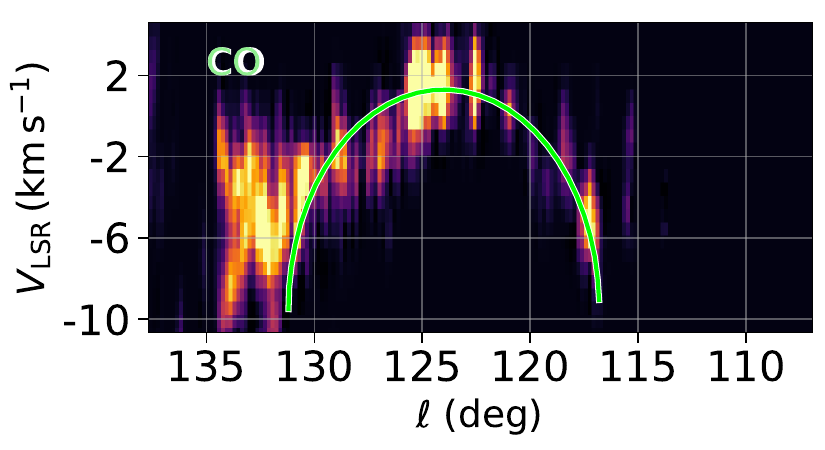}
    \includegraphics[width=0.48\linewidth, trim=.2cm 0.0cm 0.5cm .2cm, clip]{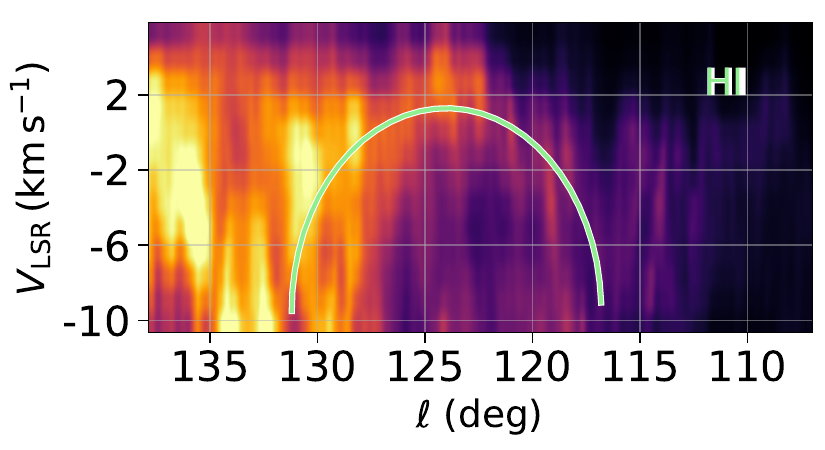}
    
    \caption{Position-velocity diagrams along the cavity (from upper-right to lower-left). Left panel: The lime solid line represents the velocity trend of CO along $\ell$ direction. Right panel: The lime solid line is the velocity trend of HI along the same direction.
    \label{fig:CO-HI_PV}}
\end{figure*} 

\subsection{Relationship with the Cepheus Flare star-formation region}
\label{sec:relationship nSF region}

The Cepheus Flare star-forming region, located 300$-$500 pc from the Sun, contains dense molecular clouds, young stellar objects (YSOs), and feedback-driven structures such as the Loop~III supernova remnant and the Cepheus Flare Shell. This region provides a unique laboratory for studying sequential star formation driven by feedback mechanisms.

In Figure~\ref{fig:Gs1-5-ext}, we show the 2D dust extinction map of the Cepheus Flare star-forming region (300$-$500~pc) based on \citet{Edenhofer2024}, on which the current distribution of the members in Groups~1$-$5, as well their median proper motions are overlapped. In the Cepheus Flare region, there are several known star-forming regions, including L1251, L1235, L1217/L1219, L1177, NGC~7023, L1148/L1158, and L1228 (see Figure~\ref{fig:Gs1-5-ext}). In these regions, \citet{Szilagyi2021MN} (thereafter S21) identified 319 YSOs with a age of 1$-$5 Myr. Among them, 149 YSOs have been included in Group~4 in this work. We re-estimate the age of these YSOs using the same method as done for Groups 1$-$5, which gives a best-fit age of $\sim$7~Myr (see Figure~\ref{fig:yso_CMD_bestfits}).

The members of Group~4 form an arc-like structure, along which the kinematics of the members exhibit expansion motion (see Figure~\ref{fig:Gs1-5-ext}). This indicates that these sources originated from the same parental cloud. However, the ages of the YSOs are much younger than those of the other members in Group~4. Isochrone fitting reveals ages of 7~Myr and 19~Myr for the YSOs and the other members in this group, respectively, suggesting a sequential star formation process occurring along the arc structure from the bottom to the top (see Figure~\ref{fig:Gs1-5-ext}). Furthermore, the age differences among Groups~1$-$5 and all known YSOs in the Cepheus Flare region hint at the presence of multiple generations of star formation in this area.

The Cepheus Flare region exhibits complex structures within the ISM, including molecular filaments, low-density cavities, and expanding shells, which are likely influenced by feedback from the massive stars in ASCC~127. In 2D dust extinction map (Figure~\ref{fig:Gs1-5-ext}), one prominent feature is a cavity centered at \(\text{($\ell$, b)} = (120^\circ, 18^\circ)\), spanning several tens of parsecs, which we interpret as a result of stellar winds and supernova explosions from earlier generations of stars in Groups~1$-$5. And this cavity is likely to overlap with the Cepheus Flare Shell (CFS), which has been studied by CO, soft X-ray and radio continuum observations \citep{Olano2006, Kun2007, Kun2009}. And there are also two filaments highlighted with lime lines, one of which is located with these YSOs associated with molecular clouds. Adjacent to the one filament is the famous Polaris region where no obvious star formation activity, and neither OB stars nor YSOs have been detected in this region\citep{Xia2022}.

While the 2D dust extinction map provides a projected view of these structures, complementary CO and HI observations are crucial for tracing their molecular and atomic components, as well as their kinematics. The prominent cavity and two filaments highlighted in the 2D dust extinction map (Figure~\ref{fig:Gs1-5-ext}) is largely aligned with the CO distribution (lime contour) in Figure~\ref{fig:Gs1-5-CO-HI}. In contrast, these features are challenging to identify in the HI distribution (grey background) because of its diffusion.

Although the morphologies of the cavity and filaments are clearly visible in the CO distribution, their kinematic signatures remain unclear. To further investigate them, we analyze the position-velocity (PV) distribution along the light-yellow line shown in Figure~\ref{fig:Gs1-5-CO-HI}. The extracted PV diagram, shown in Figure~\ref{fig:CO-HI_PV}, displays that the CO and HI velocity distributions exhibit a prominent inverted "U"-shaped pattern ranging from $-$11\ \text{to}\ 6\ \text{km/s}, particularly for CO. This suggests the presence of an expanding cavity in the region, instead of a shell structure. Using the analytical model provided by \citet{McCray1987}, the expansion time of the cavity with the radius between 50 and 90~pc is approximately 3 to 5~Myr.

Large expanding cavity are often attributed to feedback mechanisms driven by massive stars, such as stellar winds, and supernova explosions. We calculate the kinematic energy required to form such a large cavity. Our analysis  cover the region with $10^{\circ} < \ell < 130^{\circ}$, $10^{\circ} < $b$ < 40^{\circ}$, and a velocity range of $-$11\ \text{to}\ 6\ \text{km/s}. Using a distance of 300~pc, the total mass of HI and \( \text{H}_2 \) are estimated to be $\sim8\times 10^4~M_{\odot}$ and $\sim9\times 10^4~M_{\odot}$. The energy required to form the cavity can be estimated using the formula: \(
E_{\text{kin}} = \frac{1}{2} M_{\text{gas}} V_{\text{exp}}^2
\), Where $V_{\rm exp}$ is the expansion velocity of the gas and is about 10\, \text{km/s}, derived from position-velocity diagram (Figure~\ref{fig:CO-HI_PV}). The resulting $E_{\rm kin}$ is  $\sim10^{50}$\,\text{erg}. Within a relatively smaller region ($110^\circ <= \ell <= 135^\circ$, $10^\circ <= $b$ <= 20^\circ$) than in this work, \citet{Olano2006} has studied the CFS and obtained a lower gas mass ($1.3\times 10^4$ \Msun) and kinematic energy ($1.58\times 10^{48}$~erg) and proposed that the CFS could be driven by a supernova explosion. And in this region, there is also another larger supernova remnant, Loop~III \citep[e.g.,][]{Berkhuijsen1973, Spoelstra1973, Kun2008}, with a distance range of 300 to 600~pc and older than the CFS \citep{Olano2006, Kun2007, Kun2009}. Some studies suggest that the CFS is located within Loop~III or partially overlaps with it \citep{Olano2006, Kun2007}. The presence of the CFS and Loop~III, caused by at least two distinct supernova burst events, indicate there is a complex star formation history in the Cepheus Flare region.

Taking the initial mass function from \cite{Kroupa(2001)}, the most massive stars in the lifetime of Groups~2$-$5 are expected to be around 9~\(\mathrm{M_\odot}\), which could release the energy of stellar wind less than \(10^{49}\) prior to their supernova explosions \citep{Fichtner2024}, and the most massive star in the lifetime of Group~1 is likely to be around 3~\(\mathrm{M_\odot}\), releasing lower energy compared to the more massive stars in Groups~2$-$5. According to the PARSEC 1.2S model with solar metallicity, a 9 \(\mathrm{M_\odot}\) star has a lifetime of approximately \(31\)\,Myr. The classical supernova explosion could release the energy of \(10^{51}\)\,erg\citep{McCray1987, Farias2024, Fichtner2024}, comparable to the energy for forming the cavity. Considering the uncertainties on the lifetimes of massive stars and age dating in the groups, it is likely that the cavity in this region is produced by the supernova explosion from one or more massive stars in the groups. The similar cavities have been found in other star-forming regions, e.g., Mon~OB1 \citep{Zhuang2024ApJ}.

\section{Summary}
\label{sec:summary}

In this work, we used Gaia~DR3 data and the FOF clustering algorithm to re-identify the nearby young moving group ASCC~127 in the Cepheus Flare star-forming region. Combined these groups with the known YSOs, we explored the star formation history in the Cepheus Flare star-forming region. The main findings of this work are summarized as follows:

\begin{itemize}
    \item 
     We identify 3,971 members in ASCC~127, which double the known members in the literature. The moving group ASCC~127 is divided into 5 subgroups, Groups~1$-$5. Among them,  Groups~1 and 4 are firstly revealed in this work. The ages of Groups~1$-$5 range from 15 to 32~Myr. Groups~1$-$5 and the nearby YSOs in Cepheus Flare star-forming region consist of four distinct age groups: 32~Myr, 20~Myr, 15~Myr, and 7~Myr.

    \item 
    We identify a cavity with a radius of several tens of parsecs in the dust map in the Cepheus Flare. And according to the CO and HI data, this cavity is expanding at around 10\,\kms, and the kinematic energy formed this cavity is around $10^{50}$\,erg.
    
    \item
    Groups~1$-$5 and the nearby YSOs sequentially formed in multiple episode, and during the process of the first two generations, massive stars ended their lives, driving the formation of the Loop~III and the Cepheus Flare Shell.
    
\end{itemize}

A combination of studies on ISM and young stellar populations enhances our understanding of the star formation history in the Cepheus Flare region.


\setcounter{table}{1}
\setcounter{table}{1}
\begin{table*}
\scriptsize
\begin{center}
\caption{Properties of Groups~1$-$5.}
\label{properties of Gs5} 

    \begin{tabular}{ccccccccccccc}
        \hline
        \hline
        \textbf{Group} & \textbf{$\ell_0$} & \textbf{$b_0$} & \textbf{$d$} & \textbf{$X$} & \textbf{$Y$} & \textbf{$Z$} & \textbf{${\mu_{l^*}}$} & \textbf{${\mu_b}$} & \textbf{$RV_{LSR}$} & \textbf{Mass} & \textbf{Age} & \textbf{Number}\\
        &(degree)&(degree)&(pc)&(pc)&(pc)&(pc)&(\masyr)&(\masyr)&(\kms) & (\Msun)&(Myr) & \\
        (1) & (2) & (3) & (4) & (5) & (6) & (7) & (8) & (9) & (10) & (11) & (12) & (13) \\
        \hline
        1 &125.38&-0.70 &$382\pm21$    &224$\pm$16   &306$\pm$24    &-5$\pm$8     &0.8$\pm$0.7   &0.7$\pm$0.5   &$-0.7\pm4.9$  &140&\textbf{$32^{+4}_{-5}$} & \textbf{140}\\
        2 &96.14&-1.16  &$434\pm34$    &35$\pm$35    &431$\pm$33    &-9$\pm$12    &-3.7$\pm$0.9  &-0.4$\pm$0.6  &$-4.9\pm8.1$  &600&\textbf{$22^{+4}_{-1}$} & \textbf{796}\\       
        3 &114.59&-3.72 &$423\pm23$    &175$\pm$23   &381$\pm$26    &-28$\pm$14   &0.2$\pm$1.0   &-0.5$\pm$0.7   &$-5.2\pm8.6$ &700&\textbf{$15^{+1}_{-2}$} &\textbf{1194}\\        
        4 &111.60&3.80  &$371\pm23$    &138$\pm$14   &344$\pm$22    &34$\pm$40    &-1.4$\pm$0.8  &-0.7$\pm$1.0   &$-7.3\pm9.2$ &560&\textbf{$15^{+1}_{-2}$} & \textbf{864}\\
        5 &106.91&0.33  &$487\pm32$    &138$\pm$35   &463$\pm$32    &-4$\pm$19    &-1.3$\pm$0.7  &-0.7$\pm$0.6   &$-3.7\pm9.0$ &800&\textbf{$20^{+2}_{-1}$} &\textbf{977}\\
        \hline  
    \end{tabular} 
\end{center}       
\footnotesize{Column\,(1): group ID. Column\,(2)$-$(3): median Galactic coordinates of each group. Column\,(4): median distance of each group and its standard deviation. Column\,(5)$-$(7): median three-dimensional heliocentric positions of each group. Column\,(8)$-$(9): median proper motions of each group with respect to LSR and its standard deviation. Column\,(10): medain radial velocity of each group with respect to LSR and its standard deviation. Column\,(11): total mass of each group. Column\,(12): age and its uncertainty of each group. Column\,(13): total number of each group.} 
\end{table*}

\acknowledgments
We thank Xiaoying Pang for the helpful discussions. This work is supported by the National Key R\&D Program of China with grant 2023YFA1608000.
X.F. acknowledges the support of the National Natural Science Foundation of China (NSFC) No. 12203100 and the China Manned Space Project with NO.CMS-CSST-2021-A08. X.-L.W. acknowledges the support by the Science Foundation of Hebei Normal University (grant No. L2024B56) and S\&T Program of Hebei (grant No. 22567617H).

\software{{\tt Astropy} \citep{AstropyCollaboration2013,AstropyCollaboration2018}, {\tt ROCKSTAR} \citep{Behroozi(2013)}, {\tt Numpy} \citep{vanderWalt2011CSE}, {\tt Matplotlib} \citep{Hunter2007CSE}, {\tt TOPCAT} \citep{Taylor2005ASPC}}.

\appendix
\section{A comparison of membership in this work and in KC19}
\label{Appen:com}

In Figure~\ref{fig:dr3vsKC19}, we compare the distribution 5$-$dimensional phase space and CMD of the ASCC~127 in KC19 and in this work. Using the FOF algorithm, we recover all the structures in KC19 besides the one located at $-20^{\circ}<b<-13^{\circ}$. In addition to the known structures in KC19, our research has uncovered several new structural features. In total, we identified 3,971 members, increasing the membership count by a factor of 2 compared to KC19.

\renewcommand{\thefigure}{\Alph{section}.\arabic{figure}}
\setcounter{figure}{0}

\begin{figure}
    \centering
        {%
            \includegraphics[width=0.95\linewidth,trim=0.1cm 0.4cm 0cm 2.7cm, clip]{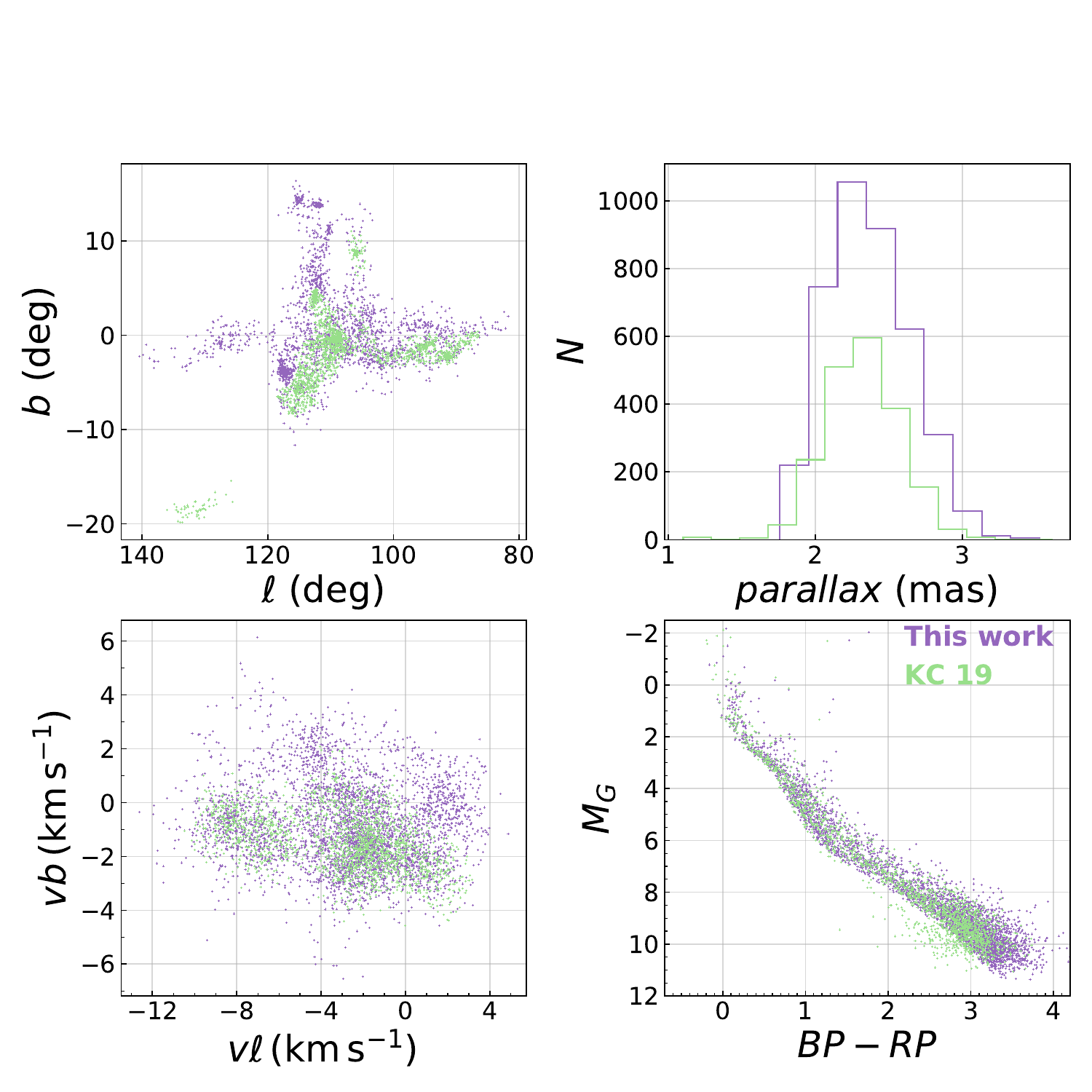}
            }

\caption{Distributions of 3,971 member candidates (purple dots) from this work and 2,124 member candidates (lime dots) from KC19 in five-parameter space (Galactocentric coordinate ($\ell$, b), parallax, tangential velocity (v$\ell$, vb)) and the CMD ($M_G$ vs. BP$-$RP). 
\label{fig:dr3vsKC19}}
\end{figure}


\end{CJK*}
\end{document}